Original article

# Improved survival of cancer patients admitted to the ICU between 2002 and 2011 at a U.S. teaching hospital

C.M. Sauer[1,2*], J. Dong[3], L.A. Celi[2], D. Ramazzotti[4]

[1] Department of Epidemiology, Harvard T.H. Chan School of Public Health, Boston, MA, United States

[2] Institute for Medical Engineering and Science, Massachusetts Institute of Technology, Cambridge, MA, United States

[3] Institute for Clinical Research and Health Policy Studies, Tufts Medical Center, Boston, MA, United States

[4] Department of Pathology, Stanford University, Stanford, CA, United States

*Corresponding author:
Dr. Leo A. Celi
Institute for Medical Engineering and Science
Massachusetts Institute of Technology
Deacon St, Cambridge, MA 02142
Tel: +1-617-710-3114
lceli@bidmc.harvard.edu

**Abstract**

**Background:** Over the past decades, both critical care and cancer care have improved substantially. Due to increased cancer-specific survival [1], we hypothesized that both the number of cancer patients admitted to the ICU and overall survival have increased since the millennium change.

**Patients and methods:** MIMIC-III, a freely accessible critical care database of Beth Israel Deaconess Medical Center, Boston, USA [2] was used to retrospectively study trends and outcomes of cancer patients admitted to the ICU between 2002 and 2011. Multiple logistic regression analysis was performed to adjust for confounders of 28-day and 1-year mortality.

**Results:** Out of 41,468 unique ICU admissions, 1,100 hemato-oncologic, 3,953 oncologic and 49 patients with both a hematological and solid malignancy were analyzed. Hematological patients had higher critical illness scores than non-cancer patients, while oncologic patients had similar APACHE-III and SOFA-scores compared to non-cancer patients. In the univariate analysis, cancer was strongly associated with mortality (OR= 2.74, 95%CI: 2.56, 2.94). Over the 10-year study period, 28-day mortality of cancer patients decreased by 30%. This trend persisted after adjustment for covariates, with cancer patients having significantly higher mortality (OR=2.63, 95%CI: 2.38, 2.88). Between 2002 and 2011, both the adjusted odds of 28-day mortality and the adjusted odds of 1-year mortality for cancer patients decreased by 6% (95%CI: 4%, 9%). Having cancer was the strongest single predictor of 1-year mortality in the multivariate model (OR=4.47, 95%CI: 4.11, 4.84).

**Conclusion:** Between 2002 and 2011, the number of cancer patients admitted to the ICU increased steadily and significantly, while longitudinal clinical severity scores remained overall unchanged. Although hematological and oncological patients had higher mortality rates than patients without cancer, both 28-day and 1-year mortality decreased significantly and faster than those of non-cancer patients over the 10-year study period.

**Key message:** Admission of cancer patients doubled between 2002-2012. Cancer patients had higher 28-day mortality rates than non-cancer patients (28.6% vs. 12.8%). While 28-day mortality rates of cancer patients decreased by 31%, clinical severity scores remained unchanged. After adjusting for confounders, there was an annual 6% decrease in the odds of dying at both 28-days and 1-year among cancer patients.

**Background**

Cancer is a global public health concern and the second leading cause of death in the United States and most developed countries [3]. Despite a slight decrease in cancer incidence over the last two decades, cancer prevalence continues to increase due to improved treatment and earlier diagnosis [4].

Historically, advanced or end-stage oncological and hematological patients were, due to their limited prognosis, frequently not referred to intensive care units. Novel and more efficient anti-neoplastic treatments have resulted in prolonged progression-free and overall survival of oncological and hemato-oncological patients [3]. Consequently, a growing number of cancer patients is at risk of admission to intensive care units [1].

Concurrently, the average prognosis of oncological patients remains less than that of the general population, thereby frequently resulting in controversy whether an ICU admission is an appropriate medical choice [1]. While several studies have described the characteristics of oncological patients admitted to ICU, only two studies have investigated trends in hematological and oncological ICU admissions in the UK [5] and hematological patients in the Netherlands [6].

In both settings, the absolute number of oncologic patients admitted to intensive care units has increased. In the Netherlands, the odds of admission increased by an annualized 6% between 2004 and 2012, while the proportion of oncological patients admitted to British ICUs remained almost unchanged between 1997 and 2013. Of note, the Acute Physiology and Chronic Health Evaluation (APACHE-II) score for hematological patients at admission were the same in the Dutch and British cohorts (24 points). At the same time, both hospital (47% vs. 54%) and ICU mortality (34% vs. 41%) were approximately 7 percentage-points lower in the Dutch ICU cohort compared to the UK cohort, possibly indicating differences in referral pattern, aggressiveness of treatment or other patient characteristics. In both cohorts, hospital and ICU mortality for hematological patients decreased, while mortality rates of oncological patients did not change. Regardless, the average prognosis of oncological patients remains less than that of the general population. For instance, hematological patients in the Netherlands had twice as high ICU mortality rates than patients without malignancy (34% vs. 17%).

While prior studies have described oncological patient characteristics and outcomes in the U.S. [7], no published longitudinal analysis for trends in ICU mortality for the US is available. In the following, we therefore aim to describe trends in admission numbers and patient characteristics for cancer patients admitted to the ICU. Furthermore, we aim to establish if 28-day and 1-year mortality changed over the 10-year period, after adjusting for covariates.

**Methods**

Patient information was extracted from a de-identified version of the MIMIC-III (v1.4) database using PostgreSQL (The PostgreSQL Global Development Group, https://www.postgresql.org). MIMIC-III is an open-access ICU database jointly administered by the Massachusetts Institute of Technology and the Beth Israel Deaconess Medical Center [2]. The database comprises over 58,000 hospital admissions for 38,645 adults and 7,875 neonates, spanning from June 2001 - October 2012. Year of admission is not available in this limited, de-identified dataset due to privacy concerns. Therefore, we obtained a supplementary dataset from the administrators. Previously established code from the MIMIC Code Repository was used to generate comorbidity scores and risk scores [8].

A waiver of consent that has previously been obtained from the Institutional Review Boards of BIDMC and MIT is applicable to these datasets.

*Cohort Selection*

All adult patients aged 18 years or older with a diagnosis of any cancer type as identified by ICD-9 codes 140-199 (solid malignancies) and 209.0-3 (neuroendocrine carcinomas) or 200-208 (hematological malignancy) were included in this study. Patients with incomplete data on admission year, length of stay or severity of illness were excluded during data analysis. Since no complete annual data for 2001 and 2012 was available, all patients admitted in this period were excluded from the base population. If patients were admitted multiple times to the ICU, only their first admission was used for this analysis.

*Covariates*

Patient data on the following variables were extracted from the database:

Age at admission, gender, race, Elixhauser Comorbidity Index scores [9], vasopressor duration, ventilation duration, OASIS scores [10], SOFA scores [11], APACHE-III scores [12], renal replacement therapy, do not resuscitate order (DNR) at admission, ICU mortality, mortality (from the Social Security death records). SOFA scores were interpreted as being low (0-5), medium (6-10) and high risk (≥11 points), respectively. Duration of ventilation and duration of vasopressor use were converted from continuous to binary variables. Race was categorized into a white and non-white category.

Cancer diagnoses were clustered according to ICD-9 codes into: Oral malignancies (1400-1499), gastrointestinal (1500-1599), respiratory and thoracic (1600-1659), Connective tissue malignancy and others (1700-1769, 17300-17399), genitourinary (1800-1899), other solid malignancies (1900-1992, 20900-20936), lymphoma (20000-20208, 20270-20288), leukemia (20310-20892), other hematologicalmalignancies (20210-20268, 20302-20382, 20290-20302) and metastatic cancer (1960-1991,20970-20975).

*Outcomes*

The primary study outcomes were (a) trends in the number and relative frequency of oncological patients admitted to the ICU and (b) all-cause mortality within 28 days and 1 year after ICU admission. Secondary outcomes were (a) hospital and ICU length of stay, (b) changes in clinical severity scores and (c) predictors of mortality.

*Statistical Analysis*

All analyses were performed using the open-access software *R* (version 3.4.2, http://www.R-project.org/) and the following packages: *tableone, ggplot2, dplyr, MIMICbook, epitools, sjplot, MASS*. For table 1, chi-square testing was performed for categorical variables (with continuity correction) and one-way ANOVA for parametric continuous variables. Multiple logistic regression was used to adjust for covariate levels in the longitudinal trend analysis for 28-day and 1-year mortality (categorical outcome). Model building was based on expert opinion including the most important confounders (C.M.S.& L.A.C.). An alpha-level of 0.05 was used as the cutoff value to reject the null hypotheses. Collinearity analyses were performed, hence only APACHE-III scores were included in the final model. We checked the final model for interaction terms, however did not find any interaction terms that were statistically significant after Bonferroni adjustment.

*Reproducibility*

Both the code for the SQL queries and data analysis will be made freely publicly available after acceptance of this article for publication. Due to privacy concerns, the re-identified data on year of admission will not be made publicly available.

**Results**

By restricting our analysis to the first ICU stay for each patient, a total unique patient cohort of 41,468 ICU admissions was obtained. Of these, 3953 had an oncologic malignancy, 1100 had a hematological malignancy and 49 patients had a diagnosis of both a hematological and solid malignancy (fig. 1).

Baseline patient characteristics differed considerably between all subgroups (table 1). Hematological patients had highest mean clinical severity scores and a higher rate of DNR orders on admission (*p<0.001*), however were less likely to be ventilated or to receive vasopressors (*p<0.001*). Furthermore, mean length of ICU and hospital stay for hematological patients was significantly longer than that of non-cancer patients (*p<0.001*). Oncological patients had patient characteristics that were more similar to those of non-cancer patient, having comparable clinical severity scores and only slightly longer length of ICU and hospital stays. Of note, oncological patients less frequently received vasopressin and ventilation.

Between 2002 and 2011, the overall number of patients admitted to the ICU increased from 3110 to 5076. Over the same period, the number of patients with cancer more than double from 285 to 678 (Suppl. table 1) and 28-day mortality of cancer patients decreased from 36.1% in 2002 to 24.8% in 2002 (*p<0.001*). Concurrently, use of vasopressors (*p=0.002*) and ventilation (*p<0.001*) decreased significantly, while clinical severity scores did not change significantly, and comorbidity scores increased from 16 to 20 (*p<0.001*).

Between 2002 and 2011, both crude 28-day and 1-year mortality rates decreased (fig. 2). The highest mortality rates were measured in the oncology group, with 27.7% and 52.6% at 28 days and 1 year, respectively. Over the study period, the highest relative decrease in the odds of death at 28 days was recorded in the hematological group (-55%), followed by the oncologic (-40%) and non-cancer patients (-20%). There was considerable variation in 28-day mortality rates depending on cancer subtype, with other oncological cancers (including neuroendocrine carcinomas) and cancer of the lip, oral cavity and pharynx having the highest mortality rate. Genitourinary and connective tissue (including breast) cancers had the lowest mortality rates of all malignancies. Mortality rates were closely associated with clinical severity scores across all cancer subtypes. Of note, patients with high SOFA-scores (≥11) and a connective tissue malignancy (including breast cancer, $N$=2) or genitourinary cancer ($N$=7) had a 100% probability of dying at 28-days. Similarly, patients with gastrointestinal cancer ($N$=49) or leukemia ($N$=23) and high SOFA-scores (≥11) had a close to 100% probability of dying within 1-year of ICU admission (Supp. Fig. 1).

Multivariate logistic regression analysis confirmed the observed trends in mortality. For the whole ICU cohort, a do-not-resuscitate order at admission, any type of cancer and year of admission were the strongest predictors of both 28-day and 1-year mortality rates after adjustment for covariates (suppl. fig. 2). A DNR order was associated with a 3.69-times higher odds (95%CI: 3.30, 4.13) of death within 28 days of admission, and a malignancy with a 2.62 (95%CI: 2.38, 2.88) times higher odds of death. Notable differences between the intensive care units exists, with the cardiac surgery recovery unit being associated with the lowest odds of dying (0.2, 95%CI: 0.17, 0.23), while admission to a trauma surgical unit was associated with a 1.25 (95%CI: 1.10, 1.43) higher odds of death. Having any malignancy was the strongest single predictor of 1-year mortality after adjustment for covariates with an odds ratio of death of 4.47 (95%CI: 4.12, 4.85). The observed protective association between admittance to the cardiac surgery recovery unit persisted (OR= 0.23, 95%CI: 0.21, 0.26). The odds of death decreased by an annualized 6% (95%CI: 5%, 7%) each year. (suppl. fig. 2)

Among cancer patients, a DNR order is the single most important factor associated with both 28-day and 1-year survival after adjustment for covariates (fig. 3). For 28-day survival, being ventilated (OR=2.12, 95%CI: 1.78, 2.52) or receiving vasopressors (OR=1.41, 95%CI: 1.16, 1.70) is strongly associated with worse outcome. Noteworthily, this association is weaker for 1-year survival, with vasopressor use (OR=1.27, 95%CI: 1.06, 1.52) being weakly associated and ventilation (OR=1.07, 95%CI: 0.91, 1.24) not being statistically significantly associated. Both Apache-III score and Elixhauser comorbidity score are important predictors of survival. Patients in the lowest Apache-III score quartile (score=31) have a 2.0-times higher odds of dying than patients in the 3rd Apache-III score quartile (score=55). Similarly, patients in the lowest Elixhauser score quartile (score=10) have a 1.7-times higher odds of dying than patients in the 3rd Elixhauser score quartile (score=28). Of note, white race is weakly associated with higher survival at 1-year (OR=0.82, 95%CI: 0.70, 0.97), but not with 28-day (OR=0.93, 95%CI: 0.77, 1.11). For both 28-day and 1-year survival, admission to a surgical, trauma surgical or cardiac surgery ICU is associated with a significantly decreased odds of death compared to admission to a medical ICU.

Over the 10-year study period, both 28-day and 1-year mortality decreased at an annualized rate of 6%. Between 2002 and 2011, the odds of dying decreased by 49.2% (28-day mortality, 95%CI: 31%, 62%) and 44.7% (1-year mortality, 95%CI: 28%, 59%).

**Discussion**

This is one of the largest studies investigating trends in patient characteristics and outcomes of oncological patients admitted to the ICU worldwide, and the first study from the U.S. More than 58,000 ICU admissions were analyzed and 41,468 patients were included in the final analysis, among which were 1100 hematological and 3953 oncological patients. Cancer patients had comparable clinical severity scores compared to non-cancer patients, yet were more likely to die within 28 days (RR=2.23) and within 1 year (RR=2.58). Mortality rates decreased significantly over the study period (-7% annual decrease) for all patients, with cancer patients having a bigger decrease in mortality rates than non-cancer patients.

Our findings confirm the previously reported decrease in mortality rates observed in the UK [5] and the Netherlands [13] since the millennium change. Interestingly, we observed unchanged clinical severity scores over time, while mortality rates decreased substantially between 2002 and 2011. While 28-day survival of oncological patients (71.1%) was comparable with 30-day survival in the Dutch cohort (73.1%), survival of hematological patients (72.3%) differed considerably from the Netherlands (55%).

As expected, clinical severity scores were strong predictors of survival, as were the presence of any malignancy, a DNR order at admission, type of ICU, use of vasopressors or mechanical ventilation and comorbidities, thereby confirming earlier studies [5, 7, 14-18]. Of note, crude 1-year mortality rates for patients with high SOFA scores and gastrointestinal, genitourinary, connective tissue cancer, leukemia, lymphoma or other solid cancers were exceeding 90%. Mortality rates differed considerably between the ICU types and are possibly explained by different reasons for admission and long-term prognosis. The notable protective effect observed for cardiac surgical patients could be explained by selection bias, since primarily patients with a favorable oncologic diagnosis might undergoing surgery. In contrast, medical ICU patients might more likely be admitted due for reasons directly related with their underlying disease or treatment complications.

Clinicians are frequently faced with the difficult decision whether patients with serious chronic conditions that limit their life expectancy should be admitted to the intensive care unit. This is particularly challenging for oncological patients, where long-term survival might more depend on their underlying malignancy than their acute health problem. Our findings together with previously published results [14, 16, 19, 20] suggests that elderly patients with high clinical severity scores, multiple comorbidities and certain cancer types do not sustain long-term survival regardless of intensive treatment. Concurrently, the presence of any malignancy alone is an insufficient predictor of both 28-day and 1-year mortality and treatment decisions should not be based on oncological prognosis alone.

Analysis of over 40,000 admitted patients, including more than 5,000 (hemato-)oncological patients allowed for precise and stratified analysis of patient characteristics, outcomes and predictive factors. Data was retrieved from clinical databases that were compiled at the time of treatment, thereby minimizing bias that may occur due to the retrospective study design. The high resolution of the clinical data is another advantage of this study. Furthermore, we had sufficient statistical power to detect annual changes in mortality of cancer patients.

Major limitations of this study are a lack of important oncological predictive factors, including TNM classification, tumor size, histopathological features and clinical aggressiveness. Furthermore, no data was available on critically ill patients who were not admitted to the ICU and abstained. Preferably, additional information on oncological treatment, treatment type and time since administration of chemotherapy should be included in the analysis. In addition, data on rare cancers (including oral cancer and respiratory cancers) was sparse, thus not allowing for a more detailed subgroup analysis. Another disadvantage is that data was obtained from a single-center teaching hospital and generalizability of findings might therefore be limited. Future analyses should preferably include larger cohorts from different hospitals across the United States and contain more oncologic outcomes measures.

Regardless of these shortcomings, the unchanged clinical severity scores over time and results of the multivariate logistic regression suggest that survival of oncologic patients has increased between 2002-2011 and is not due to confounding by other variables or selection bias. It is tempting to speculate that the absolute and relative increase in cancer patients admitted to the ICU could be interpreted as a consequence of better oncological survival, rendering more patients at risk of ICU admission. The alternative, namely that the observed increase in oncological patient admissions is due to a more lenient ICU admission policy, is not supported by unchanged clinical severity scores, increased comorbidity scores and another study showing a decrease in ICU utilization among cancer patients between 2002-2013 [7]. Interestingly, 28-day mortality for both cancer and non-cancer patients decreased approx. equally on the multiplicative scale (-6% annually). At the same time, the absolute improvement in survival is larger for cancer patients, due to an approx. three times higher mortality rate at baseline.

In this descriptive study, which should not be interpreted as a prognostic survival model, we recorded a steady and significant increase in the number and proportion of cancer patients admitted to the ICU, unchanged clinical severity scores, increased comorbidity score and a major decrease in mortality rates between 2002 and 2011. In our multivariate model for the whole ICU cohort, the presence of any malignancy was the single strongest predictor of 1-year survival of all ICU patients.

**Acknowledgements**

This manuscript has been produced as part of course HST.953 at the Massachusetts Institute of Technology, Cambridge, MA, USA. The authors gratefully acknowledge the support and feedback provided by T. Pollack, A. Johnson and J. Raffa.

**Funding**

The MIMIC database is funded by the National Institute of Health through the National Institute of Biomedical Imaging and Bioengineering [grant number R01 EB017205-01A1].

**Disclosure**

**Legend to the figures**

**Figure 1: Overview of the cohort building process.** Restriction to the years for which complete follow-up data is available decreased the cohort by 5,009 individuals. A total of 5,102 patients with cancer were identified, of which 3,953 had an ICD-9 code of a solid tumor, 1,100 of a hematological malignancy and 49 both a hematological and solid malignancy.

**Figure 2: Changes in 28-day mortality for (A) solid malignancies and (B) hematological malignancies over the 10-year period.** There are notable differences in mortality between the subgroups, with cancers of the oral cavity/lip/pharynx and other primary origin having the highest mortality rates. Genitourinary (GU) and breast/connective tissue had the lowest mortality rates. GI: gastrointestinal cancer, Oral: cancer of the oral cavity/ lip/ pharynx

**Figure 3: Results of the multivariate regression model for (A) 28-day and (B) 1-year mortality among cancer patients.** (A) DNR at admission (OR=3.4) was the strongest single predictor of 28-day mortality after adjusting for covariates. Any use of vasopressors or mechanical ventialtion was associated with worse 28-day outcomes. Year of admission was a strong predictive factor, with the odds of dying within 28-days decreasing by 7% every year. Elixhauser and Apache-III scores were signficanly associated with both 28-day and 1 year mortality, e.g. a 1 point increase in APACHE-III increasing the odds of dying within 28 days by 4%. Admission to a ICU other than the medical was associated with increased survival. (B) Notable differences between 28-day and 1-year survival are the weaker association between both vasopressor use and ventialtion with survival and a signficant assoication between white race and survival.

**Legend to the tables**

**Table 1: Overview of patient characteristics and outcomes.** Patients with cancer had higher comorbidity scores, similar clinical severity scores, on average received less invasive treatment and had higher 28-day and 1-year mortality. All differences between the subgroups where overall statistically significant with *p*-values<0.001. DNR=Do not resuscitate, SOFA: Sepsis related organ failure assessment, OASIS: Oxford acute severity of illness score, APACHE-III: Acute physiology and chronic health evaluation score. If not indicated differently, values refer to the mean (Standard Deviation).

**References**

1. Jemal A, Ward EM, Johnson CJ et al. Annual Report to the Nation on the Status of Cancer, 1975-2014, Featuring Survival. J Natl Cancer Inst 2017; 109.
2. Johnson AEW, Pollard TJ, Shen L et al. MIMIC-III, a freely accessible critical care database. Scientific Data 2016; 3: 160035.
3. Siegel RL, Miller KD, Jemal A. Cancer Statistics, 2017. CA Cancer J Clin 2017; 67: 7-30.
4. Miller KD, Siegel RL, Lin CC et al. Cancer treatment and survivorship statistics, 2016. CA Cancer J Clin 2016; 66: 271-289.
5. Ostermann M, Ferrando-Vivas P, Gore C et al. Characteristics and Outcome of Cancer Patients Admitted to the ICU in England, Wales, and Northern Ireland and National Trends Between 1997 and 2013. Crit Care Med 2017; 45: 1668-1676.
6. van Vliet M, Verburg IW, van den Boogaard M et al. Trends in admission prevalence, illness severity and survival of haematological patients treated in Dutch intensive care units. Intensive Care Med 2014; 40: 1275-1284.
7. Wallace SK, Rathi NK, Waller DK et al. Two Decades of ICU Utilization and Hospital Outcomes in a Comprehensive Cancer Center. Crit Care Med 2016; 44: 926-933.
8. Johnson AE, Stone DJ, Celi LA, Pollard TJ. The MIMIC Code Repository: enabling reproducibility in critical care research. J Am Med Inform Assoc 2018; 25: 32-39.
9. Moore BJ, White S, Washington R et al. Identifying Increased Risk of Readmission and In-hospital Mortality Using Hospital Administrative Data: The AHRQ Elixhauser Comorbidity Index. Med Care 2017; 55: 698-705.
10. Johnson AE, Kramer AA, Clifford GD. A new severity of illness scale using a subset of Acute Physiology And Chronic Health Evaluation data elements shows comparable predictive accuracy. Crit Care Med 2013; 41: 1711-1718.
11. Vincent JL, Moreno R, Takala J. The SOFA (Sepsis-related Organ Failure Assessment) score to describe organ dysfunction/failure. On behalf of the Working Group on Sepsis-Related Problems of the European Society of Intensive Care Medicine. Intensive Care Med 1996; 22.
12. Knaus WA, Wagner DP, Draper EA et al. The APACHE III prognostic system. Risk prediction of hospital mortality for critically ill hospitalized adults. Chest 1991; 100: 1619-1636.
13. Bos MM, Verburg IW, Dumaij I et al. Intensive care admission of cancer patients: a comparative analysis. Cancer Med 2015; 4: 966-976.
14. Rosolem MM, Rabello LS, Lisboa T et al. Critically ill patients with cancer and sepsis: clinical course and prognostic factors. J Crit Care 2012; 27: 301-307.
15. Azevedo LCP, Caruso P, Silva UVA et al. Outcomes for patients with cancer admitted to the ICU requiring ventilatory support: results from a prospective multicenter study. Chest 2014; 146: 257-266.
16. Soares M, Caruso P, Silva E et al. Characteristics and outcomes of patients with cancer requiring admission to intensive care units: a prospective multicenter study. Crit Care Med 2010; 38: 9-15.
17. Neuschwander A, Lemiale V, Darmon M et al. Noninvasive ventilation during acute respiratory distress syndrome in patients with cancer: Trends in use and outcome. Journal of Critical Care 2017; 38: 295-299.
18. Wigmore T, Farquhar-Smith P. Outcomes for Critically Ill Cancer Patients in the ICU: Current Trends and Prediction. Int Anesthesiol Clin 2016; 54: e62-75.
19. Soares M, Salluh JI, Toscano L. Outcomes and prognostic factors in patients with head and neck cancer and severe acute illnesses. Intensive Care Med 2007; 33.
20. Demandt AMP, Geerse DA, Janssen BJP et al. The prognostic value of a trend in modified SOFA score for patients with hematological malignancies in the intensive care unit. Eur J Haematol 2017; 99: 315-322.

**Supplementary material**

**Supplementary table 1: trends in cancer patient admission, characteristics and outcomes by year of admission.** The number of oncological cases admitted to the ICU increased by 238% over the 10-year period. Clinical severity scores remained approximately unchanged, while use of vasopressors and ventilation decreased significantly. The mean Elixhauser comorbidity score increase steadily over the study period. Both 28-day and 1-year mortality decreased over time, with the 28-day mortality (-31%) decreasing faster than 1-year mortality (-18%).

**Supplementary figure 1: (A) 28-days and (B) 1-year mortality rates by cancer type and SOFA-scores.** SOFA scores were strongly associated with mortality across all cancer subtypes. Overall, cancer patients had worse outcomes compared to non-cancer patients with similar SOFA scores. Patients with high SOFA-scores (≥11) and a connective tissue malignancy (including breast cancer, *N*=2) and genitourinary cancer (*N*=7) had a 100% probability of dying at 28-days. Similarly, patients with gastrointestinal cancer (*N*=49) and leukemia (*N*=23) and high SOFA-scores (≥11) had a close to 100% probability of dying at 1-year.

**Supplementary figure 2: Results of the multivariate model for (A) 28-day and (B) 1-year mortality for the whole cohort.** (A) After adjusting for other covariates, DNR at admission (OR=3.69) was the strongest predictor of 28-day mortality, followed by any use of mechanical ventilation and presence of any malignancy. Year of admission was a strong protective factor, with mortality rates decreaseing by 6% every year. With each year increase in age at admission, the odds of dying within 28 days and 1 year increased by 2%. A one point increase in APACHE-III scores was closer asscoiated with a higher increase in the odds of dying within 28 days (4%) than 1 year (2%).

| Variables | Without cancer | Hematologic Malignancy | Solid malignancy | Hematologic and solid cancer |
|---|---|---|---|---|
| Number of cases | 36,366 | 1,100 | 3,953 | 49 |
| Age at admission, years | 63.8 (17.8) | 65.3 (16.4) | 65.8 (13. 6) | 69.9 (12.7) |
| DNR at admission (%) | 1819 (6.0) | 72 (8.2) | 297 (9.5) | 3 (7.7) |
| White race (%) | 26074 (71.7) | 851 (77.4) | 2979 (75.4) | 39 (79.6) |
| Length of stay ICU, days | 4.59 (6.09) | 5.38 (7.15) | 4.19 (5.30) | 4.00 (3.34) |
| Length of stay Hospital, days | 9.3 (8.7) | 14.7 (15.2) | 10.1 (8.3) | 11.0 (10.0) |
| Use of ventilation (%) | 18181 (50.0) | 470 (42.7) | 1699 (43.0) | 20 (40.8) |
| Use of vasopressors (%) | 12428 (34.2) | 369 (33.5) | 990 (25.0) | 11 (22.4) |
| Elixhauser score | 8.8 (11.1) | 17.6 (11.8) | 19.4 (13.0) | 23.9 (12.4) |
| SOFA score | 3.3 (2.5) | 4.6 (3.0) | 3.3 (2.7) | 3.4 (2.9) |
| APACHE-III score | 42.0 (19.4) | 50.4 (22.1) | 44.5 (20.3) | 46.7 (23.3) |
| OASIS score | 31.4 (9.0) | 32.5 (9.0) | 31.2 (9.5) | 31.2 (9.9) |
| 28-day Mortality (%) | 4640 (12.8) | 305 (27.7) | 1143 (28.9) | 13 (26.5) |
| 1-year Mortality (%) | 8378 (23.0) | 579 (52.6) | 2420 (61.2) | 29 (59.2) |

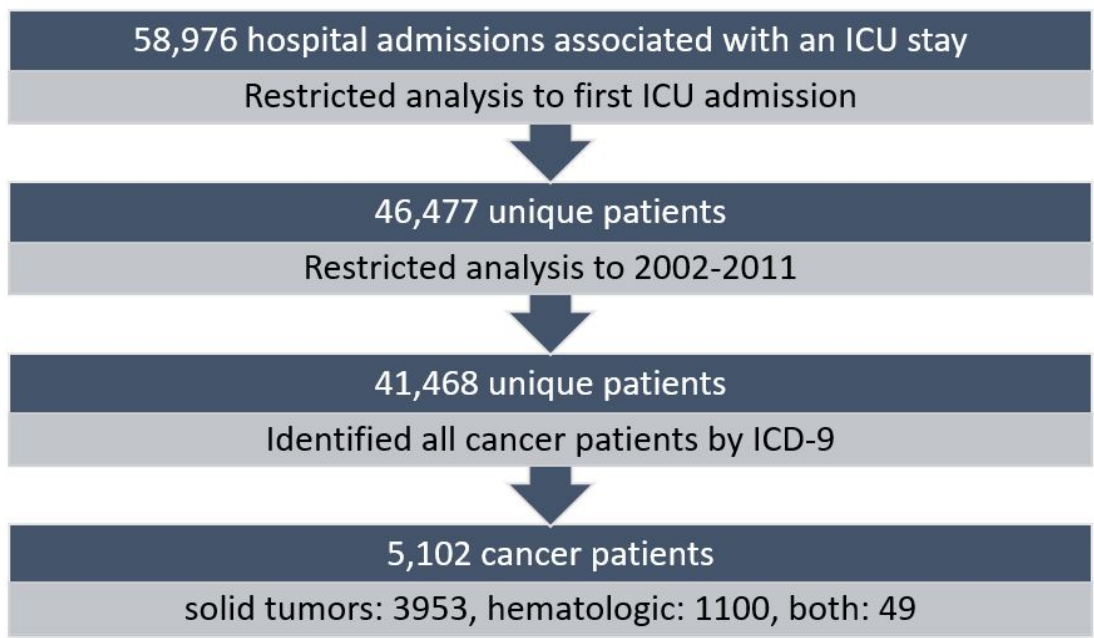

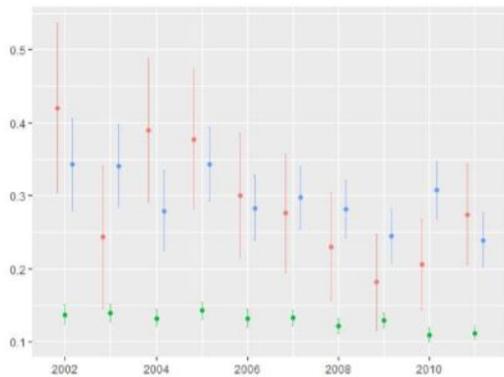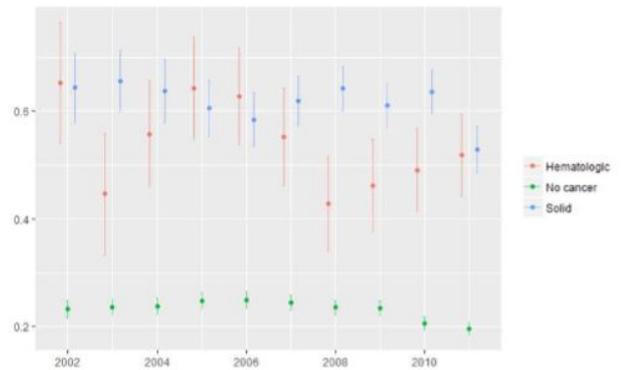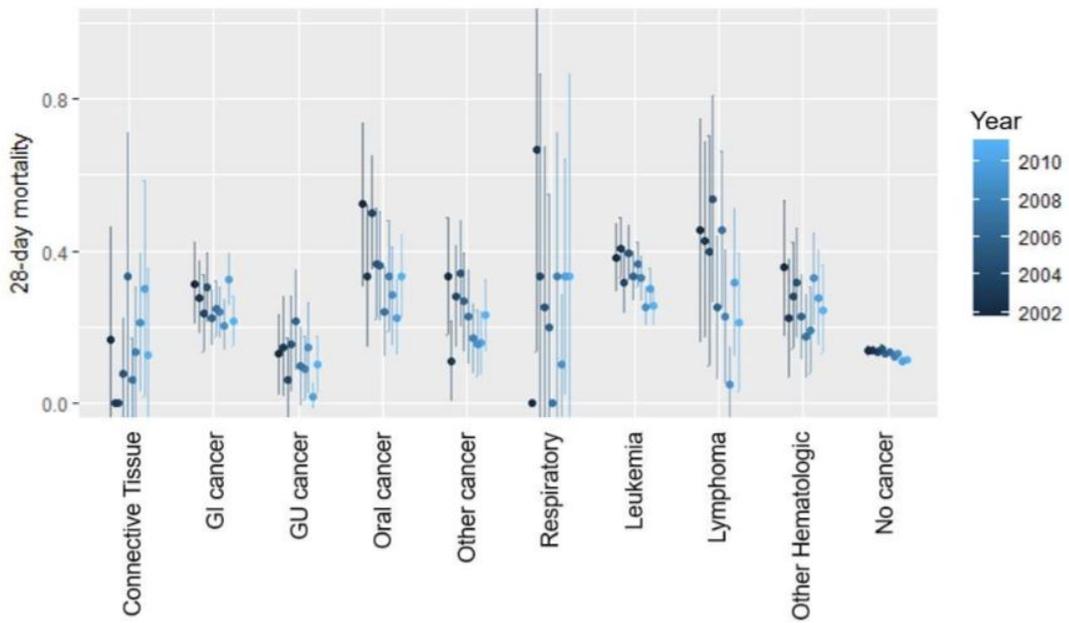

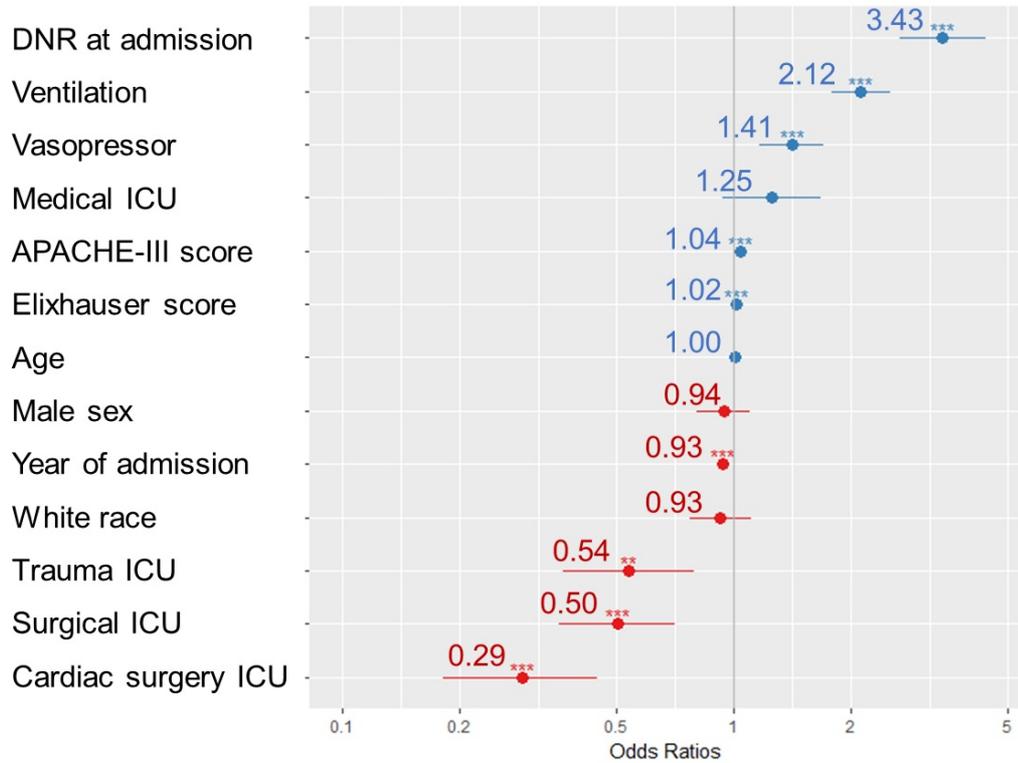

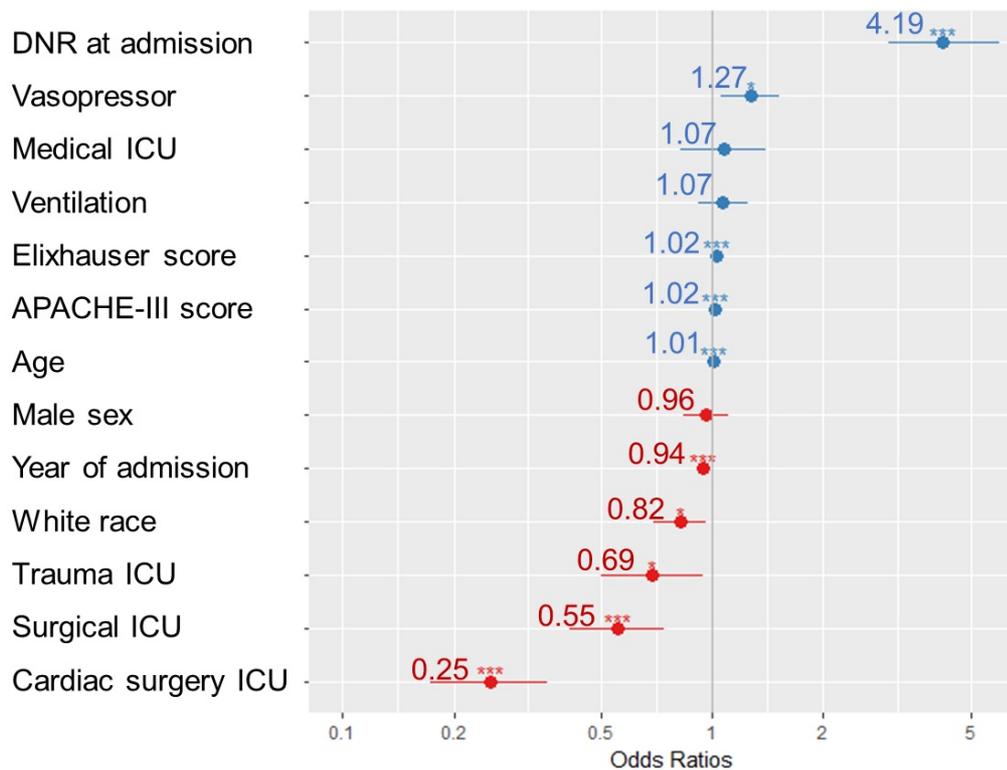

| Variables | 2002 | 2003 | 2004 | 2005 | 2006 | 2007 | 2008 | 2009 | 2010 | 2011 | p-value |
|---|---|---|---|---|---|---|---|---|---|---|---|
| Number of Patients | 285 | 341 | 349 | 433 | 495 | 560 | 630 | 643 | 688 | 678 | 0.012 |
| Age at admission, year | 65.58 (14.06) | 64.02 (14.92) | 64.23 (15.46) | 64.28 (13.92) | 65.23 (14.10) | 66.60 (13.85) | 66.29 (14.11) | 65.92 (13.94) | 66.33 (14.04) | 66.61 (14.12) | 0.012 |
| White race (%) | 169 (59.3) | 247 (72.4) | 252 (72.2) | 323 (74.6) | 385 (77.8) | 433 (77.3) | 495 (78.6) | 526 (81.8) | 528 (76.7) | 511 (75.4) | <0.001 |
| Use of vasopressors (%) | 89 (31.2) | 106 (31.1) | 115 (33.0) | 130 (30.0) | 138 (27.9) | 145 (25.9) | 165 (26.2) | 154 (24.0) | 177 (25.7) | 151 (22.3) | 0.002 |
| Use of ventilation (%) | 133 (46.7) | 171 (50.1) | 164 (47.0) | 224 (51.7) | 218 (44.0) | 233 (41.6) | 243 (38.6) | 280 (43.5) | 288 (41.9) | 235 (34.7) | <0.001 |
| Solid Malignancy (%) | 216 (75.8) | 267 (78.3) | 254 (72.8) | 335 (77.4) | 385 (77.8) | 444 (79.3) | 504 (80.0) | 511 (79.5) | 523 (76.0) | 514 (75.8) | 0.226 |
| Hematologic Malignancy (%) | 69 (24.2) | 74 (21.7) | 95 (27.2) | 98 (22.6) | 110 (22.2) | 116 (20.7) | 126 (20.0) | 132 (20.5) | 165 (24.0) | 164 (24.2) | |
| Metastatic Cancer (%) | 79 (27.7) | 137 (40.2) | 109 (31.2) | 149 (34.4) | 157 (31.7) | 200 (35.7) | 216 (34.3) | 230 (35.8) | 227 (33.0) | 218 (32.2) | 0.065 |
| Renal replacement therapy (%) | 14 (4.9) | 15 (4.4) | 6 (1.7) | 20 (4.6) | 22 (4.4) | 19 (3.4) | 33 (5.2) | 23 (3.6) | 31 (4.5) | 35 (5.2) | 0.298 |
| Elixhauser Score | 15.87 (10.99) | 17.28 (11.69) | 17.50 (12.11) | 19.18 (12.75) | 18.13 (12.55) | 19.69 (12.74) | 19.45 (12.81) | 19.96 (12.88) | 20.07 (13.19) | 19.90 (13.71) | <0.001 |
| SOFA score | 3.40 (2.95) | 3.52 (2.71) | 3.57 (2.79) | 3.87 (2.93) | 3.56 (2.81) | 3.48 (2.81) | 3.76 (2.87) | 3.44 (2.48) | 3.74 (2.80) | 3.46 (2.77) | 0.108 |
| APACHE-III score | 46.52 (23.77) | 44.58 (21.21) | 46.03 (20.45) | 45.89 (21.49) | 45.60 (21.87) | 45.23 (19.66) | 47.21 (22.80) | 45.21 (19.29) | 47.15 (21.19) | 44.42 (18.52) | 0.258 |
| OASIS score | 31.49 (10.05) | 30.69 (9.82) | 31.44 (9.80) | 31.53 (9.42) | 31.46 (9.38) | 31.38 (8.94) | 31.83 (9.89) | 31.54 (9.12) | 32.15 (8.94) | 30.63 (9.37) | 0.188 |
| ICU length of stay, days | 5.34 (8.72) | 5.32 (5.83) | 5.40 (6.70) | 4.97 (6.78) | 4.89 (6.86) | 4.03 (4.53) | 4.20 (5.62) | 3.97 (4.07) | 4.30 (5.71) | 3.67 (4.05) | <0.001 |
| Hospital length of stay, days | 11.80 (11.13) | 11.31 (10.48) | 12.92 (12.08) | 11.28 (9.58) | 11.91 (10.88) | 10.90 (9.80) | 10.57 (9.25) | 10.71 (12.04) | 10.98 (10.53) | 9.91 (8.56) | 0.001 |
| 28-day Mortality | 103 (36.1) | 109 (32.0) | 108 (30.9) | 152 (35.1) | 142 (28.7) | 164 (29.3) | 171 (27.1) | 149 (23.2) | 195 (28.3) | 168 (24.8) | <0.001 |
| 1-year Mortality | 184 (64.6) | 208 (61.0) | 215 (61.6) | 266 (61.4) | 294 (59.4) | 339 (60.5) | 378 (60.0) | 373 (58.0) | 414 (60.2) | 357 (52.7) | 0.027 |

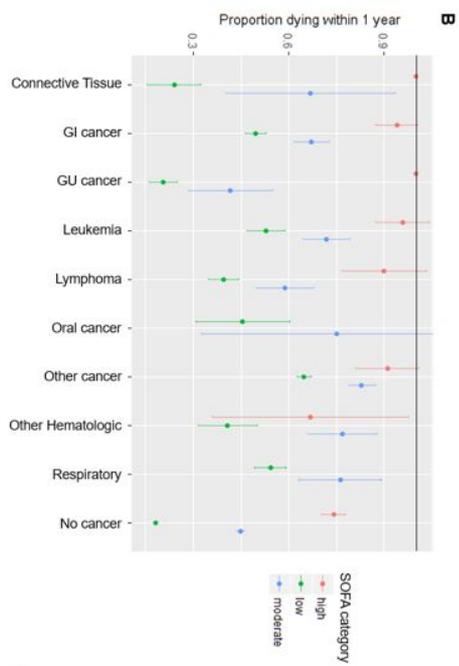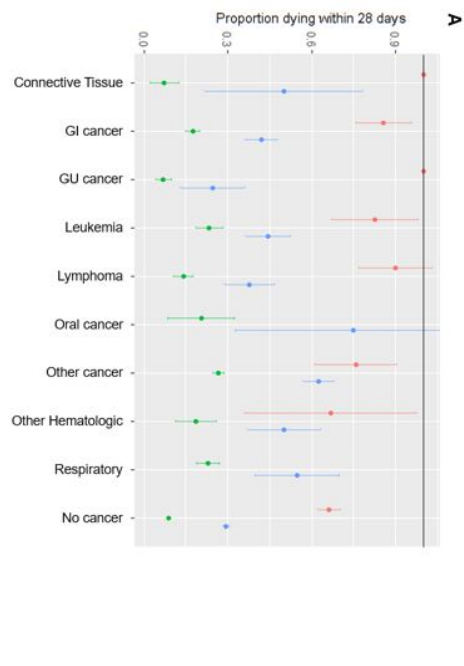

A

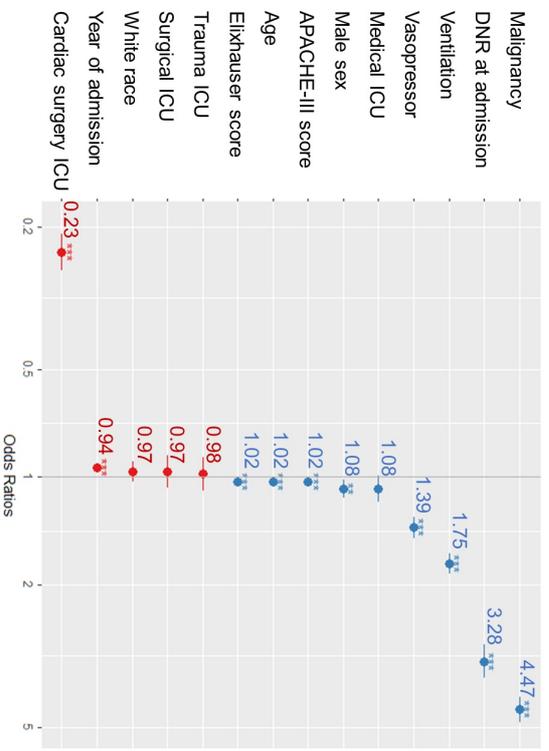

B

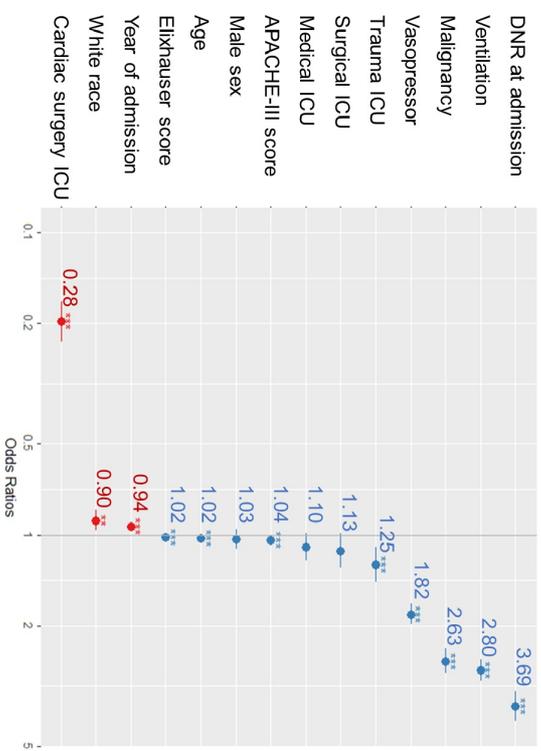